 \documentclass[useAMS]{mn2e}
\usepackage{graphicx,graphics,amsmath}
\usepackage{longtable}
\title{Discovery of a 0.02 Hz QPO feature in the Transient X-ray Pulsar KS 1947+300}

\author[Marykutty James, Biswajit Paul, Jincy Devasia and Kavila Indulekha]{Marykutty James$^{1,2}$\thanks{E-mail: marykuttykjames@yahoo.co.in}, Biswajit Paul$^{2}$, Jincy Devasia$^{1,2}$ and Kavila Indulekha$^{1}$\\
$^{1}$School of Pure and Applied Physics, Mahatma Gandhi University, Priyadarshini Hills P.O, Kottayam-686560, Kerala, India\\
$^{2}$Raman Research Institute, Sadashivnagar, C. V. Raman Avenue, Bangalore 560080, India}
\begin{document}

 \pagerange{\pageref{firstpage}--\pageref{lastpage}} \pubyear{2009}
\maketitle
 \label{firstpage}

\begin{abstract}
We report the discovery of Quasi Periodic Oscillations (QPO) at 0.02 Hz in a
transient high mass X-ray binary pulsar KS 1947+300 using {\em RXTE}-PCA.
The QPOs were detected during May-June 2001,
at the end of a long outburst. This is the 9th transient accretion
powered high magnetic field X-ray pulsar in which QPOs have been detected and
the QPO frequency of this source is lowest in this class of sources. 
The unusual feature of this source is that though the outburst lasted for more
than 100 days, the QPOs were detected only during the last few days of the
outburst when the X-ray intensity had decayed to 1.6\% of the peak intensity.
The rms value of the QPO is large, $\sim15.4\pm1.0\%$  with a slight positive
correlation with energy. The detection of QPOs and strong pulsations at a low
luminosity level suggests that the magnetic field strength of the neutron
star is not as high as was predicted earlier on the basis of a correlation
between the spin-up torque and the X-ray luminosity.
\end{abstract}


\section{INTRODUCTION}

KS 1947+300 is an accretion powered X-ray pulsar which was first detected in
June 1989 in observations made with the TTM coded-mask imaging spectrometer
aboard the {\em Kvant} module of the {\em Mir Space Station}
(Skinner 1989). The 2-27 keV flux was $70\pm10$ mCrab. In
two months the flux from the source faded by a factor of seven. Its X-ray
spectrum could be described by a power law with photon index
$\Gamma = 1.72\pm0.31$ and
a hydrogen column density $N_{H} = (3.4\pm3.0) \times 10^{22}$ (Borozdin et al.
1990). The coordinates of the source were determined to be : R.A.$ = 19^{h}47^{m}35^{s}.2$, Dec.= $+30^{0} 04'47''$
(Eq. 1950.0). In April 1994, the Burst and {\em Burst and Transient Source Experiment} (BATSE) aboard the {\em Compton Gamma Ray Observatory
(CGRO)} detected 18.7 s pulsations from an X-ray source within a few degrees
of KS 1947+300. The newly detected source GRO J1948+32 was later found to be
same as KS 1947+300 (Swank and Morgan 2000). The optical counterpart is a
V=$14.2$ B0 Ve star with  moderate reddening that indicates the distance to
the system to be about 10 kpc (Negueruela et al. 2003). \\

One large outburst and several smaller outbursts of KS 1947+300 have been
observed with the {\em Rossi X-ray Timing Explorer (RXTE)}-ASM. The first outburst
of this source was detected by {\em RXTE}-ASM in October 2000 (Levine and
Corbet 2000). Following this the intensity declined, but the source became highly active again
 in November 2000. The outburst reached its peak in February 2001
and slowly declined till June 2001. Based on
data acquired with {\em RXTE-PCA} during the 2000-2001 outburst, Galloway et
al.  (2004) determined the orbital parameters of the binary: the orbital period
$P_{orb}=40.415\pm0.010$ d, the projected semi major axis $a_{x}sini =
137\pm3$ lt-sec and eccentricity $e = 0.033\pm0.013$. Glitches are mainly
observed in Anomalous X-ray Pulsars (AXP) and radio pulsars. But the {\em RXTE}
analysis of this source revealed an increase in pulse frequency at an unusually
high rate giving evidence for the first time for a glitch  in an accretion
powered pulsar (Galloway 2004). A broad band (0.1-100.0 keV) study of this
source was first carried out with {\em BeppoSAX}. This revealed that the energy
spectrum has three components - a Comptonized component, a 0.6 keV blackbody
component, and a narrow and weak iron emission line at 6.7 keV (Naik et al.
2006). The absorption column density measured towards this source is low 
(4.0-5.0$\times$ 10$^{21}$ atoms cm$^{-2}$).\\

Quasi Periodic Oscillations (QPOs) in X-ray binaries are generally thought
to be related to the rotation of the inner accretion disk (Paul \& Rao 1998).
 Any inhomogeneous
matter distribution or blobs of material in the inner disk may result in QPOs
in the power spectrum. In the case of accretion powered X-ray pulsars this
 gives useful information about the interaction between the accretion
disk and the neutron star magnetosphere. Black hole X-ray binaries and low magnetic field
neutron stars show QPOs over a wide range of frequency from a few Hz to a few
hundred Hz. High magnetic field neutron star systems show only low frequency
QPOs, in the range 10 mHz upto about 1 Hz. We have investigated the timing
properties of the transient X-ray pulsar KS 1947+300 using observations made
with the RXTE-PCA and report here the discovery of a transient QPO feature in
this source.

\section{OBSERVATIONS AND DATA}

KS 1947+300 was observed extensively by {\em RXTE} during 2000-2002. {\em RXTE}
carries three X-ray astronomy instruments. The All Sky Monitor (ASM) is
sensitive to X-ray photons between 1.5 and 12 keV. The Proportional Counter Array
(PCA) consists of five xenon proportional counter detectors, sensitive in the
energy range of 2-60 keV with an effective area of 6500 cm$^{2}$ at 6 keV. The High Energy
Timing Experiment (HEXTE) operates in the energy band of 15-250 keV. 
 
In Figure 1 we show the one day averaged ASM light curve of this source during
2000-2002 in 1.5-12 keV energy band. In this period the source showed one
large outburst that lasted for about four months, and several smaller
outbursts. The first PCA observation of KS 1947+300 was performed in 2000
November 21. Subsequently, observations were made every 2 to 3 days until June
18 around when the main outburst ended. Some more PCA observations were
carried out again in 2002, around a smaller outburst. A background
subtracted lightcurve of the
source made with data from one of the RXTE PCA detectors, PCU2 is shown in
Figure 2. We have used all PCA observations of this source available in the
archive.\\

Any contamination from other sources are negligible because there is no known
hard X-ray source within 
2 degree from KS 1947+300.
A total of 674 ks of useful data was obtained with the PCA from
135 observations. The details of the observations are
given in Table~\ref{log-pca}. 

\begin{figure}
\includegraphics[height=3.3in, angle=-90]{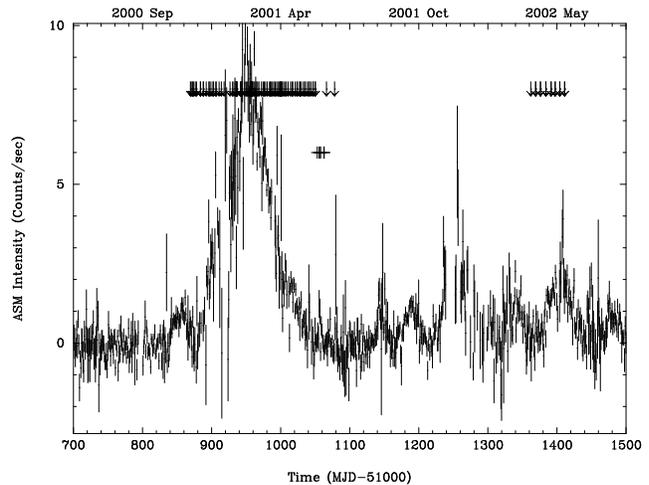}
\caption{One day averaged {\em RXTE}-ASM light curve of KS 1947+300 during 2000-2002 in the 1.5-12 keV energy band. The vertical arrows indicate the times of the {\em RXTE}/PCA observations and the '+' signs
indicate the period when the QPOs are detected.}
\label{ASM}
\end{figure}

\begin{figure}
\includegraphics[height=3.3in, angle=-90]{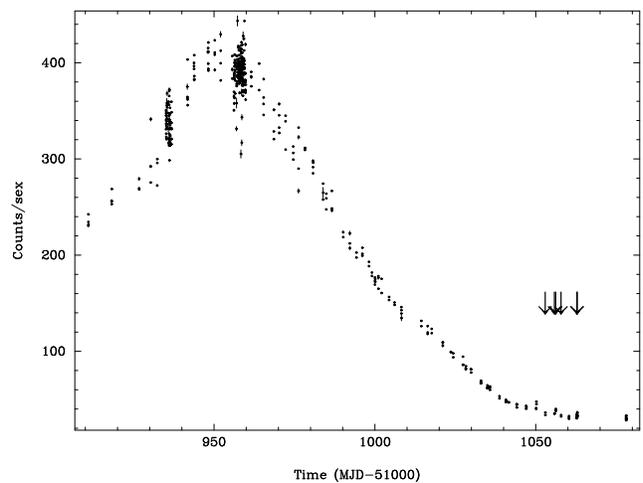}
\caption{ Background subtracted light curve of KS 1947+300
obtained with {\em RXTE}-PCU2 during the 2001 outburst with a bin size of 800 s. The vertical arrows
indicate the times of the QPO detection.}
\label{pcu2}
\end{figure}

{
\begin {table}
 \caption{List of PCA Observations}
~\\
\begin {tabular}{|c|c|c|c|}
 \hline
Year&Obs Ids&No. of &Total Durations\\
    &       &Pointings&(ks)\\
\hline
2000&P50425&19&111.83\\
 \hline
2001&P50068&30&280.1\\
    &P50425&18&73.7\\
    &P60402&53&162.8\\
\hline
2002&P70404&15&46.3\\
\hline
\end{tabular}
\label{log-pca}
\end{table}
}

\subsection{Timing and Spectral Analysis}

\begin{figure}
\includegraphics[height=3.3in, angle=-90]{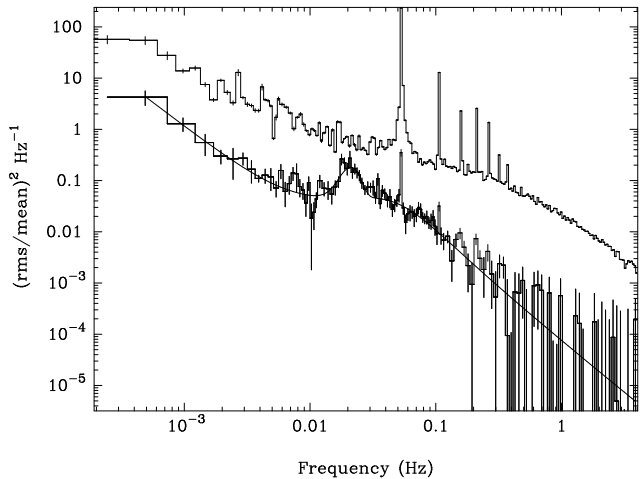}
\caption{\textbf The power spectrum of KS1947+300 obtained during the peak of
the outburst (top curve) and during the QPO detection period
are shown here. The top curve has been multiplied by an arbitrary factor.}
\label{PDS}
\end{figure}

  We extracted 2-60 keV light curves from the PCA observations using the
Standard-1 mode data which has a time resolution of 0.125 s. To search for
QPOs we created Power Density Spectra (PDS) using the FTOOL {\em powspec}
for small data segments of duration 2048 s. PDS from 5-10 consecutive
segments were averaged to improve the detectability of any QPO like feature.
The PDS were normalized such that their integral gives the squared
rms fractional variability and the white noise level was subtracted.
A narrow peak at around $0.055$ Hz corresponding to the spin frequency of
the pulsar and several harmonics are seen in all the PDS generated from
the data of all the PCA observations in 2000-2002.
In addition to the peaks due to the pulsations, a QPO feature is seen
at $0.0215\pm0.0007$ Hz in the 2001 May and June observations.
The period during which the QPOs have been detected is also marked in
Figure~\ref{ASM} and Figure~\ref{pcu2}. Two PDS, one obtained during the peak of the
outburst and one with the QPO feature are shown in Figure~\ref{PDS}. At first,
we fitted the second PDS with two continuum components, a power-law and
a Lorentzian and obtained a $\chi^{2}$ of 178 for 110 degrees of freedom.
The fit clearly showed the need for a third component around 0.2 Hz. With
the addition of a third component, a gaussian, the $\chi^{2}$ was reduced
to 119 for 107 degrees of freedom. This improvement in $\chi^{2}$ by 59
for the addition of one component is very significant. The ratio of the
amplitude and the uncertainty of the gaussian component indicates a
6 $\sigma$ detection of the QPO feature. The peaks corresponding
to the pulsations are shown in Figure~\ref{PDS}, but these were not included
while fitting the PDS continuum. From the fitted power spectrum shown in
Figure~\ref{PDS}, and after correcting for the background count rate, we calculated the
rms value of the QPO to be quite large, $15.4\pm1.0\%$. The Quality
factor Q = $\nu/FWHM$ of the 0.02 Hz QPO is 3.6, comparable to the quality
factor of 4-10 in other HMXB pulsars.
 In PDS created with data segments of shorter duration the 
QPO feature has poor signal to noise ratio. However, by looking at the
PDS averaged over smaller segments we have verified that the broad nature of
the QPO seen in Figure~\ref{PDS} is intrinsic and it is not due to averaging of a
narrow QPO feature with a variable frequency.\\

\begin{figure}
\includegraphics[height=3.3in, angle=-90]{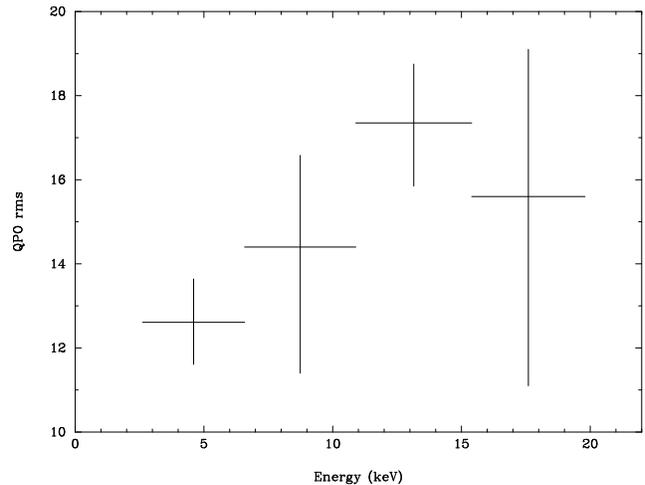}
\caption{RMS fluctuation in the 0.02 Hz QPO is shown here as a function of energy.}
\label{rms-e}
\end{figure}

We have also calculated the energy dependence of the QPO feature. We created power
spectra in different energy bands of 2.6-6.6, 6.6-11.0, 11.0-15.4 and 15.4-19.8
keV using event data from the PCA. There is marginal evidence for an increase
in the rms value with energy and is shown in Figure~\ref{rms-e}. The PCA
detectors have small effective area at higher energy and we did not detect
the QPO feature above 20 keV.\\

To measure the X-ray flux during the period when QPOs are detected, we have
generated an X-ray spectrum in 129 binned channels of the PCA observations.
A model for the background spectrum was generated using {\em pcabackest} with
appropriate background model provided by XTE guest Observer facility (GOF).
We employed the caldb version 1.0.2 and  the appropriate response matrix
was generated for this observation using pcarsp version 11.7.\\

\begin{figure}
\includegraphics[height=3.7in, angle=-90]{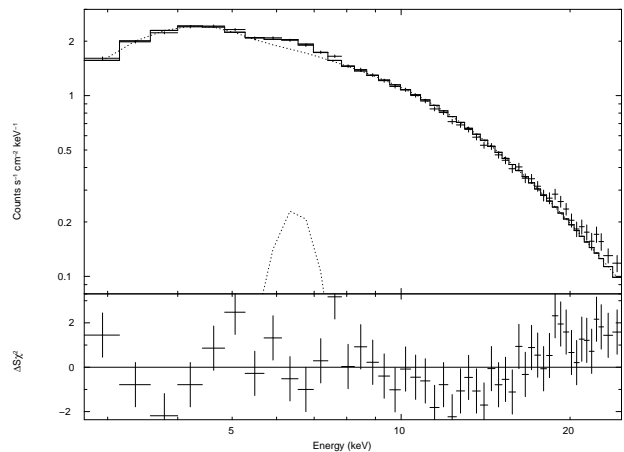}
\caption{The X-ray spectrum of KS 1947+300 is fitted with power law, absorption and a narrow emission line.}
\label{spectrum}
\end{figure}

The spectrum is fitted with an absorbed power law and an iron emission line.
The best fitted spectrum has a photon index of  1.44$\pm$0.02 and an emission line with
equivalent width of about 200 $\pm$37eV.
The $\chi_{r}^{2}$ value for the spectrum is 1.7 for 44 degrees of freedom.
The flux in the 2-20 keV band during the time of the QPO detection
 is 5.0$\times$10$^{-11}$erg cm$^{2}$ s$^{-1}$. The spectrum is shown in Figure~\ref{spectrum} along with the
best fit model and the residuals. The parameters of the best fitted model are
given in Table~\ref{spec-parameter}.\\

{
\begin {table}
\centering
  \caption{Best fit Spectral Parameters of KS 1947+300}
~\\
\begin {tabular}{|c|c|}
 \hline
Parameter&Value\\
\hline
$N_{H} (10^{22} cm^{-2})$&2.43$\pm$0.4\\
Photon index&1.44$\pm$0.02\\
norm&0.0049$\pm$0.0006\\
Gaussian line (keV)&6.50$\pm$0.12\\
Equivalent width (keV)&0.2$\pm$0.037\\
Gaussian norm&0.000055 $\pm$0.000012\\
\hline
\end{tabular}
\label{spec-parameter}
\end{table}

\section{Discussion}

 We reported the detection of a low
frequency QPO feature in the HMXB pulsar KS 1947+300 at $\nu_{QPO} = 0.02$ Hz.
The power density spectra of accretion powered pulsars consist of narrow peaks
corresponding to the spin frequency of the pulsar and its harmonics,
 accompanied by aperiodic variabilities like broad bumps.
 QPOs have been detected in 16 accretion powered high magnetic
field pulsars. These include many HMXBs and a few LMXBs; both transient  and persistent. 
The list of transient and persistent sources with QPO features are given in Table 3.
There are only three LMXB sources in which QPOs have been detected. 
Table 3 summarizes the QPO detections in high magnetic field accreting pulsars,
giving spin frequency $\nu_{s}$, the observed QPO frequency $\nu_{QPO}$, and
ratio of the two. The frequency of the QPO feature, detected in KS 1947+300 is
lowest among the transient X-ray pulsars.\\

Various models have been proposed to explain the sub-Hz QPO features in high mass X-ray binaries.
Among the most popular models are the Magnetospheric Beat
Frequency Model (MBFM; Alpar and Shaham 1985) and the Keplerian Frequency
Model (KFM; van der Klis et al. 1987). In the KFM the
QPOs arise from the modulation of the X-rays by inhomogeneities in the
accretion disk, at the Keplerian frequency. In the MBFM, the QPO frequency is
the difference between the spin frequency and the Keplerian frequency of the
inner edge of the accretion disk $\nu_{QPO}= \nu_{k}- \nu_{s}$ (Shibazaki
\& Lamb 1987). Mass flow to the neutron star is expected to be stopped at magnetospheric 
boundary by centrifugal inhibition of accretion if the Keplerian frequency at
the magnetospheric boundary is less than the 
neutron star spin frequency (Stella et al. 1986).
Among the sources listed in Table~\ref{QPO-list}, the KFM is not applicable
for some of the sources including KS 1947+300 as the observed QPO frequency
is smaller than the pulse frequency. At larger mass accretion rate,
the accretion disk is expected to extend closer to the neutron star, and
therefore, in either model, a positive correlation is expected between the QPO centroid frequency 
and the X-ray intensity (Finger 1998). In some sources, the QPO
frequency is found to be quite constant with more than a factor of 10 variation
in X-ray luminosity. In some of these sources, it is believed that the inner
disk origin may not be applicable while in some other sources there are a
variety of other reasons for a lack of correlation (Raichur \& Paul 2008a,b).\\
  
A positive energy dependence of the QPO rms, similar to that seen in
KS 1947+300 is found in some sources (XTE J1858+034, Paul \& Rao 1998). This
favours the MBFM over the KFM,
while the QPO rms is independent of the energy in the persistent X-ray pulsar
Cen X-3 (Raichur \& Paul 2008a).

With some exceptions (4U 1626-67; Kaur et al. 2008 but see also Jain et al.
2009), the QPOs are transient phenomena in all types of pulsars. The persistent
X-ray pulsars do not show persistent QPOs (Raichur \& Paul 2008a). In the
transient X-ray pulsars that show QPOs, the feature is not detected in all
the outbursts. For example the recent outbursts of the source A 0535+262 
does not show the flux dependent QPO features that was observed with BATSE
during a large outburst in 1994 (Finger et al. 1996). However, if QPOs are
present during an outburst, it is usually present throughout the outburst
(Finger et al. 1996, Mukherjee et al. 2006). In the case of KS 1947+300,
the QPO feature was not seen in 2000 and 2002 data. It appeared very
near the end of the 2001 outburst during MJD 52052-52062.
We also analysed the 2-20 keV spectrum during the peak of the outburst
and the flux measured} is $3\times$10$^{-9}$ erg cm$^{-2}$ s$^{-1}$. The QPO feature is seen to
be present when the X-ray flux had dropped to 1.6\% of this peak value.  \\

Assuming MBFM for KS 1947+300, the Keplerian frequency of the inner disk
corresponding to a 0.02 Hz oscillation is
$\nu_{k}= 0.02+0.053 = 0.073$.
Thus the radius of the inner accretion disk can be estimated as
\textbf{\begin{equation}
 R_{BFM} = \left({GM}\over{4 \pi^{2} (\nu_{QPO}+\nu_{s})^{2}}\right)^{1/3}
\end{equation}}
The $R_{BFM}$ obtained for a mass of  $1.4 M_{\odot}$ is $9.6 \times 10^{3}$ km.                                           
The X-ray flux of KS 1947+300 in the 2-20 keV band is $5.0 \times 10^{-11}$
erg cm$^{-2}$ $s^{-1}$ during the time of QPO detection. This flux amounts to an
X-ray luminosity of $0.6\times 10^{36}$ erg s$^{-1}$ for a source distance
of 10 kpc. Using a correlation between the spin-up torque and the broad band
X-ray luminosity of this source a very strong magnetic field of
 $B = 2.5 \times 10^{13}$ G was inferred for the neutron star (Tsygankov
\& Lutovinov 2005). Using the above, and canonical values for the stellar
radius and mass of 10 km and 1.4 $M_{\odot}$ respectively, the size of the
magnetosphere for a dipole field configuration can be estimated as (Ghosh \& Lamb 1991)

\begin{eqnarray}\nonumber
R_{m} = 2.4 \times 10^3 ({M}/{1.4 M_{\odot}})^{1/7}\times 
({B}/{2.5 \times 10^{12} G})^{4/7}\times \\
({R}/{10^6 cm}) ^{10/7}\times 
({L_{x}}/{3.1 \times 10^{37} ergs~s^{-1}})^{-2/7} km
\end{eqnarray}


The magnetospheric radius is obtained as $2.76\times 10^4$ km.

Although the magnetic field value is rather uncertain, such a strong
magnetic field implies a magnetospheric radius larger by a factor of 2.8
compared to the inner disk radius at
which the QPOs are likely to be produced in the MBFM and by a factor of 2
compared to the corotation radius. This is true even if we assume a 
50\% bolometric correction, 12\% larger distance (Kiziloglu et al. 2007) and the 
neutron star mass and radius larger by 20\% (Lattimer \& Prakash 2007).
 If the magnetic field is so strong, the
X-ray flux value during the QPO detection indicates that the source should
be in propeller regime. The presence of QPOs and the strong pulsations
at the low flux value thus suggests that the magnetic field may not be as
high as was suggested by Tsygankov \& Lutovinov (2005). On the other hand,
for an X-ray luminosity of $0.6\times 10^{36}$ erg s$^{-1}$ d$_{10 kpc}^{2}$
if the observed QPOs are produced due to inhomogeneities near the
magnetospheric radius, it implies a magnetic field strength of about
$B = 4 \times 10^{12}$ d$_{10 kpc}$ G for the neutron star. The X-ray
spectrum of KS 1947+300, however, is devoid of any cyclotron absorption
line corresponding to such a magnetic field strength (Naik et al. 2006).

\section{Acknowledgements} 
We thank an anonymous referee for many suggestions that helped us to improve the paper.
 This research has made use of data obtained
through the High Energy Astrophysics Science Archive Research
Center Online Service, provided by the NASA/Goddard Space Flight
Center.

\clearpage

{
\begin {table}
\begin{minipage}{160mm}
 \caption{List of QPO sources}
~\\
\begin {tabular}{|c|c|c|c|c|c|}
 \hline
Source&Type&$\nu_{s}$&$\nu_{QPO}$&$\nu_{QPO}$/$\nu_{s}$&Reference\footnotemark{}\\
    &   &(mHz)&(mHz)&(mHz)&\\
\hline
Transient pulsars& & & &\\
\hline
KS 1947+300&{HMXB}&53&20&0.38&This work\\
SAX J2103.5+4545&{HMXB}&2.79&44&15.77&1\\
A0535+26&HMXB&9.7&50&5.15&2\\
V0332+53&HMXB&229&51&0.223&3\\
4U 0115+63&HMXB&277&62&0.224&4\\
XTE J1858+034&HMXB&4.53&110&24.3&5\\
EXO 2030+375&HMXB&24&200&8.33&6\\
XTE J0111.2-7317&HMXB&32&1270&39.68&7\\
GRO J1744-28&LMXB&2100&20000&9.52&8\\
\hline
Persistent pulsars& & & &\\
\hline
SMC X-1&HMXB&1410&10&0.0071&9\\
Her X-1&LMXB&806&13&0.016&10\\
LMC X-4&HMXB&74&0.65-20&0.0087-0.27&11\\
Cen X-3&HMXB&207&35&0.17&12\\
4U 1626-67&LMXB&130&48&0.37&13\\
X Per&HMXB&1.2&54&45&14\\ 
4U 1907+09&HMXB&2.27&69&30.4&15\\
\hline
\end{tabular}
\label{QPO-list}
\vskip 0.5cm
\footnotetext{}{References:
(1) Inam et al. 2004;
(2) Finger et al. 1996;
(3) Takeshima et al. 1994;
(4) Soong \& Swank 1989;
(5) Paul \& Rao, 1998;
(6) Angelini et al. 1989;
(7) Kaur et al. 2007;
(8) Zhang et al. 1996.
(9) Angelini et al. 1991;
(10) Moon et al. 2001b;
(11) Moon et al. 2001a;
(12) Takeshima et al. 1991;
(13) Shinoda et al. 1990;
(14) Takeshima 1997;
(15) In't Zand et al. 1998; }
\end{minipage}
\end{table}
}

\end{document}